\documentclass[12 pt, amssymb,prb,showpacs]{revtex4}
\usepackage{epsfig}
\usepackage{dcolumn}
\usepackage{amsmath}
\usepackage[a4paper,dvips]{geometry}
\begin{document}

\title{\bf Hall magnetoresistivity  response under Microwave excitation revisited}
\author{Jes\'us I\~narrea }
 \affiliation {$^1$Escuela Polit\'ecnica
Superior,Universidad Carlos III,Leganes,Madrid,Spain  and \\
$^2$Instituto de Ciencia de Materiales, CSIC,
Cantoblanco,Madrid,28049,Spain.}
\date{\today}
\begin{abstract}
We theoretically analyzed the microwave-induced modification of the
Hall magnetoresistivity in high mobility two-dimensional electron
systems. These systems present diagonal magnetoresistivity
oscillations and zero-resistance states when are subjected to
microwave radiation. The most surprising modification of the Hall
magnetoresistivity is a periodic reduction which correlates with a
periodic increase in the diagonal resistivity. We present a model
that explains the experimental results considering that radiation
affects directly only the diagonal resistivity and the observed Hall
resistivity changes are coming from the tensor relationship between
both of them.
\end{abstract}
\maketitle
\newpage

 Microwave-Induced Resistivity Oscillations
(MIRO)\cite{zudov,studenikin}  and Zero Resistance States
(ZRS)\cite{mani,zudov2} are some of the most striking physical
phenomena recently discovered in the field of condensed matter
physics. MIRO and ZRS are produced in the diagonal resistivity
($\rho_{xx}$) of a two-dimensional electrons system (2DES) when is
subjected simultaneously to a static and moderate magnetic field
($B$) and Microwave (MW) radiation. A very intense activity is being
developed on this topic and recent
experimental\cite{mani,zudov2,willett,mani2,studenikin,zudov3,yang,mani3,zudov4,smet},
and
theoretical\cite{girvin,dietel,lei,ryzhii,rivera,shi,andreev,ina,mirlin,auer}
contributions are being published in a continuous basis. Some
theoretical contributions have been presented giving explanation for
some of the emerging experimental
outcomes\cite{ina2,lei2,ina3,ahn,ina4,lei3,ina5}. However no much
attention has been paid to the study of the influence of MW
radiation on the Hall magnetoresistivity ($\rho_{xy}$). On the one
hand, experimental results has been obtained\cite{mani4,studenikin2}
showing that $\rho_{xy}$ indeed present some remarkable features as,
unexpected oscillations which are in anti-phase with $\rho_{xx}$
oscillations, the experimental curve presents an average negative
slope vs $B$, and finally, the MW-induced correction to $\rho_{xy}$
tends to vanish in ZRS regions. On the other hand, almost no
theoretical effort, with some exception\cite{ryzhii2}, has been
presented to date to explain this striking results on $\rho_{xy}$.

In this letter we propose a theoretical explanation on the
$\rho_{xy}$ response obtained under MW excitation. We consider that
radiation affects directly just $\rho_{xx}$ and the observed
$\rho_{xy}$ changes are coming only from the tensor relationship
between both of them. Thus, following our approach, MW-induced
$\rho_{xx}$ response needs first to be studied. In a recently
presented model by the author\cite{ina,ina5}, it was demonstrated
that when a 2DES is subjected to a perpendicular $B$ and MW
radiation, the electronic Larmor orbit centers oscillate back and
forth in the $x$ direction (current direction) with the same
frequency as the MW field: {\it MW driven Larmor orbits}. An
important extension of this model \cite{ina,ina5} is presented here
which allows to consider linear polarized MW radiation with the
electric field vector oriented in any direction of the $x-y$
two-dimensional (2D) plane. For that purpose, symmetric gauge has
been introduced to represent the vector potential of $B$:
$\overrightarrow{A_{B}}=-\frac{1}{2}\overrightarrow{r}\times
\overrightarrow{B}$. We first obtain the $exact$ expression of the
electronic wave vector for a 2DES in a perpendicular $B$ and MW
radiation\cite{ina,ina5}:
\begin{eqnarray}
&&\Psi(x,y,t)=\phi_{N}\left[(x-X-a(t)),(y-b(t)),t\right]\nonumber  \\
&&\times  exp \frac{i}{\hbar} \left[m^{*}\left(\frac{d
a(t)}{dt}x+\frac{d b(t)}{dt}y\right)+
\frac{m^{*}w_{c}[b(t)x-a(t)y]}{2}-\int_{0}^{t} {\it L}
dt'\right]\sum_{p=-\infty}^{\infty} J_{p}(A_{N}) e^{ipwt}\nonumber\\
\end{eqnarray}
where  $\phi_{N}$ are analytical solutions for the Schr\"{o}dinger
equation with a 2D parabolic confinement, known as Fock-Darwin
states\cite{fock}. Fock-Darwin states converge to a Landau level
spectrum when $B$ is large enough or it is the only source of
confinement (present case). $X$ is the center of the orbit for the
electron spiral motion, $w$ is the frequency of the MW field and
$w_{c}$ the cyclotron frequency. ${L}$ is the classical lagrangian,
and $J_{p}$ are Bessel functions\cite{ina,ina5}. $a(t)$ for the
x-coordinate and $b(t)$ for the y-coordinate, are the classical
solutions for a driven 2D harmonic oscillator whose expressions
depend on the direction of the linearly polarized MW field. If the
polarization is aligned with the current (x-direction), i.e., the
harmonic driving force is acting only in the x-direction, then the
expressions are:
\begin{eqnarray}
a(t)&=&\frac{e
E_{o}}{m^{*}\sqrt{(w_{c}^{2}-w^{2})^{2}+\gamma^{4}}}\cos
wt=A_{1}\cos wt \nonumber\\
 b(t)&=&\frac{e
E_{o}w_{c}}{m^{*}\sqrt{w^{2}(w_{c}^{2}-w^{2})^{2}+w_{c}^{2}\gamma^{4}}}\sin
wt=A_{2}\sin wt
\end{eqnarray}
$\gamma$ is a material and sample dependent damping parameter which
affects dramatically the MW-driven electronic orbits movement and
which has been introduced phenomenologically. Along with this
movement there occur interactions between electrons and lattice ions
yielding acoustic phonons and producing a damping effect in the
electronic motion. In ref [21], we introduced a microscopical model
to calculate $\gamma$ obtaining a numerical value of $\gamma\simeq
10^{12}s^{-1}$ for GaAs . $E_{o}$ is the amplitude of the MW field.
In the case of a MW field aligned with the y-direction (the harmonic
driving force is acting in the y-direction): $a(t)=A_{2}\cos wt$ and
$ b(t)=A_{1}\sin wt$.

The first important result is that, apart from phase factors, the
$exact$ wave function, $\Psi(x,y,t)$, is the same as a Fock-Darwin
state\cite{fock} where the center of the orbits performs and
elliptical motion given by
$\frac{a^{2}}{A_{1}^{2}}+\frac{b^{2}}{A_{2}^{2}}=1$ for x-linearly
polarized MW. In the case of MW polarized in the y direction, the
elliptical motion is according to
$\frac{a^{2}}{A_{2}^{2}}+\frac{b^{2}}{A_{1}^{2}}=1$. When $w_{c}=w$
the motion becomes circular for both cases. Another important
outcome from our model is that irrespective of the linear
polarization direction, the elliptical motion for the orbit center
is reflected in the current direction as oscillatory with the same
frequency as the MW field. This MW-induced oscillatory motion will
affect dramatically the way in which electrons in their orbits
interacts with scatterers compared to the dark case. Thus, we
introduce the scattering suffered by the electrons due to charged
impurities randomly distributed in the sample. Following the model
described in [19], firstly we calculate the electron-charged
impurity scattering rate $1/\tau$ (being $\tau$ the scattering
time). Secondly we find the average effective distance advanced by
the electron in every scattering jump, $\Delta X^{MW}=\Delta X^{0}+
A_{1,(2)}\cos w\tau$, where $\Delta X^{0}$ is the effective distance
advanced when there is no MW field present\cite{ina}. The magnitude
$A_{1,(2)}$ is the amplitude of the oscillatory motion in the
current direction, depending on the orientation of the MW linear
polarization: subindex 1 corresponds to $x$ and 2 to $y$. Finally
the diagonal or longitudinal conductivity $\sigma_{xx}^{MW}$ can be
calculated: $\sigma_{xx}^{MW}\propto \int dE \frac{\Delta
X^{MW}}{\tau}(f_{i}-f_{f})$,  being $f_{i}$ and $f_{f}$ the
corresponding electron distribution functions for the initial and
final states respectively and $E$ energy. To obtain $\rho_{xx}$ we
use the well-known tensor relation
$\rho_{xx}=\frac{\sigma_{xx}}{\sigma_{xx}^{2}+\sigma_{xy}^{2}}$,
where $\sigma_{xy}\simeq\frac{n_{i}e}{B}$, being $n_{i}$ the
impurity density.

In Fig.1, we represent in two panels calculated $\rho_{xx}$,
$\rho_{xy}$ and $\Delta\rho_{xy}=\rho_{xy}^{MW}-\rho_{xy}^{dark}$ vs
$B$ for $w=50 GHz$. In the top panel, $\rho_{xx}$ with and without
MW on the left x-axis and $\rho_{xy}$ on the right one. It can be
observed clearly the typical MIRO and ZRS in $\rho_{xx}$ and the
linear dependence of $\rho_{xy}$ with $B$. In the bottom panel we
present $\rho_{xx}$ with MW on the left x-axis and $\Delta\rho_{xy}$
on the right x-axis. For all cases x-linearly polarized MW has been
used. There appears to be an oscillatory variation in
$\Delta\rho_{xy}$ where a reduction in magnitude correlates with an
increase  in $\rho_{xx}$. It is remarkably also that the calculated
$\Delta\rho_{xy}$  curve presents an average negative slope vs $B$.
Finally it is demonstrated that the MW-induced correction to the
Hall resistivity, disappears as $\rho_{xx}\rightarrow 0$. In other
words, the plot illustrates similar features as in
experiments\cite{mani4}.

In Fig. 2, we represent the same as in Fig.1, with the exception of
$\rho_{xy}$ vs $B$ which is not represented, and for $w=100 GHz$. A
similar behavior is obtained. This striking behavior for
$\Delta\rho_{xy}$ can be readily explain observing carefully its
developed expression:
\begin{eqnarray}
\Delta\rho_{xy}&=&\rho_{xy}^{MW}-\rho_{xy}^{dark}=\frac{\sigma_{xy}}{(\sigma_{xx}^{MW})^{2}+
\sigma_{xy}^{2}}-\frac{\sigma_{xy}}{(\sigma_{xx}^{dark})^{2}+\sigma_{xy}^{2}}\nonumber\\
&\simeq&
\left[\frac{B}{n_{i}e}\right]^{3}[(\sigma_{xx}^{dark})^{2}-(\sigma_{xx}^{MW})^{2}]
\end{eqnarray}
where we have taken into account that
$(\sigma_{xx}^{dark},\sigma_{xx}^{MW}) \ll \sigma_{xy}$. Following
our model, we have considered for MW and dark cases the same
expression for $\sigma_{xy}$: $\sigma_{xy}\simeq\frac{n_{i}e}{B}$.
Thus, we propose that the full MW effect on $\rho_{xy}$ is coming
only from $\sigma_{xx}^{MW}$ through the tensor relationship and
that $\sigma_{xy}$ is unaffected by the MW field. Considering that,
in the range of moderate $B$ we are working with,
$\sigma_{xx}^{dark}$ is practically constant (see, for instance, top
panel of Fig. 1), the important features of $\Delta\rho_{xy}$ are
going to depend mainly on the term $[-(\sigma_{xx}^{MW})^{2}] $.
This would explain that the corresponding oscillations of
$\sigma_{xx}^{MW}$ would be reflected as anti-phase oscillations of
$\Delta\rho_{xy}$: MW-induced increases (decreases) in
$\sigma_{xx}^{MW}$ will produce decreases (increases) in $\Delta
\rho_{xy}$. In the same way, $\Delta \rho_{xy}$ behaves as an
oscillating curve around an average straight line with negative
slope as a function of $B$. Remember that for moderate values of
$B$, $-B^{3} \rightarrow -B$, i.e., a straight line of negative
slope. This can be clearly observed in Figs. 1 and 2, (see
dashed-dotted line in bottom panels, blue color on line). Finally,
when $\sigma_{xx}^{MW}\rightarrow 0$, (ZRS region) we will obtain
$\Delta \rho_{xy} \rightarrow
\left[\frac{B}{n_{i}e}\right]^{3}[(\sigma_{xx}^{dark})^{2}]$. We
have calculated, using experimental parameters\cite{mani4}, that
$\sigma_{xx}^{dark}$ has an average value of $\sigma_{xx}^{dark}
\simeq 5 \times 10^{-6} \Omega^{-1}$. Then for an average $B$, we
can estimate that $\Delta \rho_{xy}\simeq 0.02 \Omega$, which is
very small. Similar behavior for all $B$ range has been found.
Therefore $\Delta \rho_{xy}$ obtains a very small value in ZRS
regions. Thus, it appears as if $\Delta \rho_{xy}$ would tend to
zero. Again this is in good agreement with experiments\cite{mani4}.

This work has been supported by the MCYT (Spain) under grant
MAT2005-06444, by the Ram\'on y Cajal program and by the EU Human
Potential Programme: HPRN-CT-2000-00144.

\newpage
\clearpage

Figure 1 caption: Top panel: on the  left x-axis $\rho_{xx}$ with MW
(single line, black color online) and without MW (dotted line, blue
color online), and on the right x-axis $\rho_{xy}$ with MW (dashed
line, red color on line). Bottom panel: $\rho_{xx}$ with MW (single
line, black color on line) on the left x-axis and
$\Delta\rho_{xy}=\rho_{xy}(MW)-\rho_{xy}(dark)$, (dotted line, red
color online) on the right x-axis. All as a function of $B$ and for
$w=50GHz$. MW is x-linearly polarized for all cases. T=1K.
\newline

Figure 2 caption: Same as in Fig.1, (except $\rho_{xy}$ vs $B$) and
for a MW frequency of $100 GHz$. We obtain a similar qualitatively
behavior of the different magnitudes versus $B$.
\newline

\begin{figure}
\centering\epsfxsize=3.5in \epsfysize=3.5in \epsffile{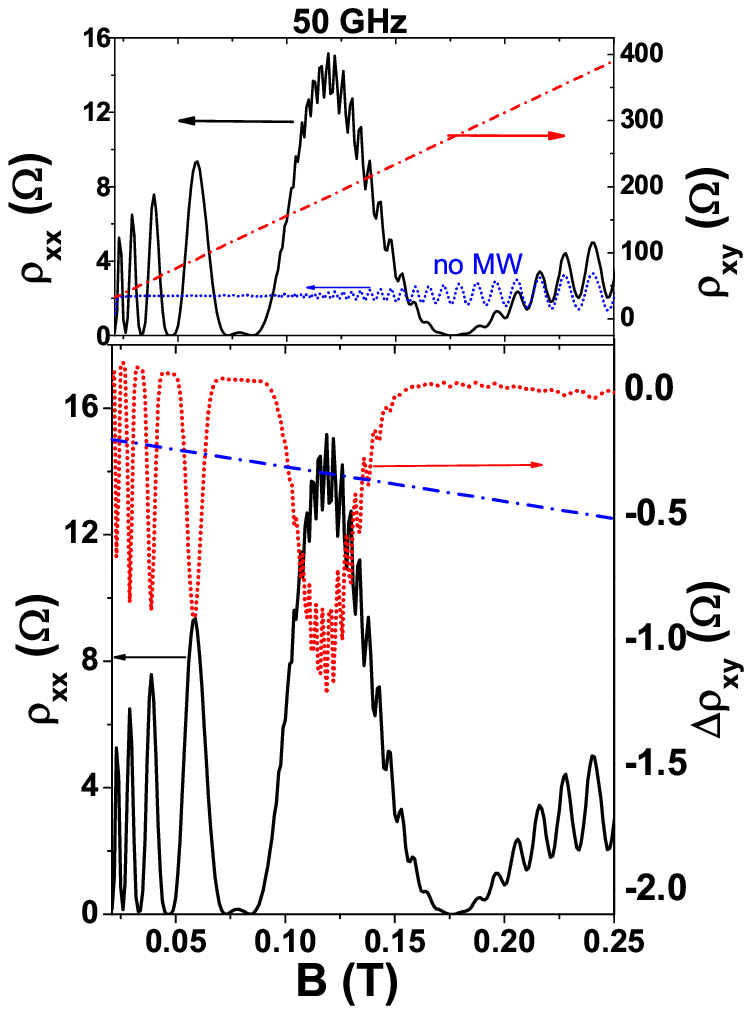}
\caption{}
\end{figure}
\begin{figure}
\centering\epsfxsize=3.5in \epsfysize=3.5in \epsffile{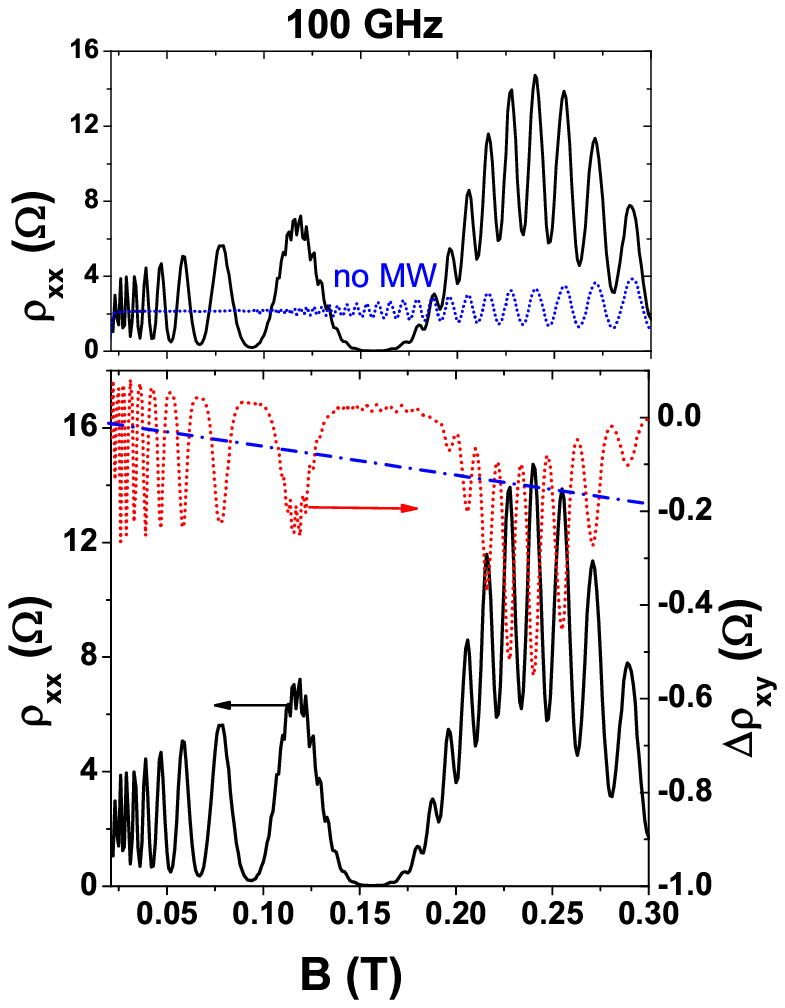}
\caption{}
\end{figure}

\end{document}